\documentclass[aps,prl,twocolumn,groupedaddress]{revtex4}

\usepackage{natbib}
\usepackage{graphicx}
\usepackage{amsmath,amssymb}
\usepackage{graphicx}
\usepackage{txfonts}

\usepackage{ulem} 

\usepackage{natbib}
\newcommand{\vect}[1]{{\boldsymbol{#1}}}

\newcommand{\bnabla}{\boldsymbol{\nabla}}

\begin{document}
\title {Transition from weak to strong cascade in MHD turbulence}
\author{Andrea~Verdini}
\email[]{verdini@oma.be}
\affiliation{Solar-Terrestrial Center of Excellence - SIDC, Royal
Observatory of Belgium, Bruxelles}
\author{Roland~Grappin}
\email[]{Roland.Grappin@obspm.fr}
\affiliation{LUTH, Observatoire de Paris, CNRS, Universit\'e Paris-Diderot, 
and LPP, Ecole Polytechnique, Palaiseau
}

 \date{\today}
 
\begin{abstract}
The transition from weak to strong turbulence when passing from large to small
scales in magnetohydrodynamic (MHD) turbulence with guide field is a
cornerstone of anisotropic turbulence theory. We present the first check of
this transition, using the Shell-RMHD which combines a shell model of
perpendicular nonlinear coupling and linear propagation along the guide field.
This model allows us to reach Reynolds numbers around $10^6$.
We obtain surprisingly good agreement with the theoretical predictions, with a reduced perpendicular energy spectrum scaling as $k_\bot^{-2}$ at large scales and as $k_\bot^{-5/3}$ at small scales, where critical balance between nonlinear and propagation time is reached. 
However, even in the strong regime, a high level of excitation is found in the weak coupling region of Fourier space, which is due to the rich frequency spectrum of large eddies. 
A corollary is that the reduced parallel spectral slope is not a definite test of the spectral anisotropy, contrary to standard belief.
\end{abstract}

   \maketitle
\textit{Introduction.}---Plasma turbulence plays an important role in solar corona, solar wind, fusion devices and interstellar medium. The knowledge of turbulent energy spectra is a basic step to solving problems like cosmic ray transport, the turbulent dynamo and solar corona or solar wind heating.
In this letter, we focus on incompressible magnetohydrodynamics (MHD) turbulence with strong guide field. 
A first theory has been first proposed in the limit of small relative magnetic fluctuation $b/B_0$, assuming isotropy of the cascade \cite{Iroshnikov:1964vb,Kraichnan:1965gi}.
This theory is based on the weak coupling of Alfv\'en waves, 
the Alfv\'en decorrelation time $t_a$ decreasing more rapidly than the nonlinear time $t_{NL}$ with wave number.
However, numerical and experimental evidence point to an anisotropic cascade, mainly in the directions perpendicular to the guide field
\citep{1981PhFl...24..825M,1983JPlPh..29..525S,Oughton:1994ue,2003PhRvE..67f6302M,2008PhRvL.101q5005H}.
This prompted \citep{1995ApJ...438..763G} to include the anisotropy within the definition of the turbulence strength as  $\chi=t_a/t_{NL} \simeq k_\bot b/(k_\parallel B_0)$ where the Alfv\'en and nonlinear times now involve respectively the parallel and perpendicular components of wave vectors. 
They thus proposed that the weak anisotropic cascade proceeds from large scales with $\chi \ll 1$ toward smaller perpendicular scales with the parallel scales remaining fixed, so that $\chi$ would increase and reach unity at some scale. The cascade would then be weak at large scales and strong at smaller scales where time scales would remain equal $\chi=1$ (critical balance or CB).
This change of regime should be characterized by a clear-cut change of spectral scaling with wave number $k$, from $k^{-2}$ to $k^{-5/3}$.

The status of the weak and strong cascades are not identical: while an analytical approach of the weak cascade is possible \cite{2000JPlPh..63..447G,2012PhRvE..85c6406S}, the strong cascade theory remains a phenomenology. 
The ansatz on weak/strong transition is a fundamental assumption in theories of anisotropic MHD turbulence 
\cite{Howes_al_2011}, but
it seems to contradict numerical results of Reduced MHD (RMHD) simulations.
In RMHD nonlinear couplings along the guide field are suppressed, which is a 
valid approximation when the guide field is strong enough
\citep{1992JPlPh..48...85Z,1993noma.book.....B}.
When $\chi_0$ was decreased starting from $1$, the spectral slope was found to vary smoothly, 
between $-5/3$ and $-3$ (with volume forcing, \cite{2008ApJ...672L..61P}), or between $-3/2$ and $-2$ (with boundary forcing, \cite{Dmitruk_al_2003,2007ApJ...657L..47R}).
In no case was the spectrum found to exhibit the predicted break between two inertial ranges.
The $-3/2$ spectral slope (instead of the $-5/3$ value)  was ascribed to a systematic weakening of interactions due to a local dominance of one Els\"asser species over the other \citep{Boldyrev:2006iu}.
These results indicate that the 
concepts of strong and weak cascades might not be as robust as generally believed, although the low Reynolds numbers reached in direct numerical simulations of RMHD might be the cause of the discrepancy.

We thus revisit this issue here, using the shell model for RMHD
\citep{Nigro_al_2004,Buchlin_Velli_2007},
in which the perpendicular coupling terms are simplified compared to RMHD, thus
allowing us to reach Reynolds numbers $Re \simeq 10^6$. In this model,
the duality between the perpendicular spectrum and the parallel space is
retained, allowing us to test the transition from weak
to strong cascade.
We consider here Shell-RMHD with volume
forcing and periodic boundaries, varying the parameter $\chi_0$ in order to compare with \cite{2008ApJ...672L..61P}.
We will first obtain the coexistence of the strong ($-5/3$) or weak ($-2$) slopes along the inertial range when forcing large eddies with $\chi_0 < 1$, in complete agreement with CB theory. Accordingly, the eddy correlation time will be found to agree either with the nonlinear time (in the strong coupling range) or with the Alfv\'en propagation time (in the weak coupling range). 
We will also show that high frequency fluctuations of the large eddies induce
an unexpected excitation level in the high $k_\parallel$ region of the spectrum.

\textit{Equations.}---The RMHD equations with mean field $B_0$ parallel to $Ox$ axis read for the
variables $\vect{z}^\pm=\vect{u}\pm \vect{b}$, with $\nabla . z^\pm = 0$
\begin{equation}
\frac{\partial \vect{z}^\pm_{\perp}}{\partial t}\mp B_0 \frac{\partial
\vect{z}^\pm_{\perp}}{\partial x} 
=-(\vect{z}^\mp_{\perp} \cdot \bnabla \vect{z}^\pm_{\perp})
- \frac{1}{\rho_0} \bnabla_{\perp} (p^T)
+\nu \bnabla ^2 _{\perp} \vect{z}^\pm_{\perp},
\label{s1}
\end{equation}
where viscosity and resistivity ($\nu$) are assumed equal. 
The shell-RMHD is obtained from RMHD by replacing the perpendicular nonlinear
couplings at each point $x$ by a dynamical system defined in Fourier space. 
The perpendicular Fourier plane is paved with $N+1$ concentric shells, the model retaining one wave number $k_n$ per shell, and one complex scalar mode
$z^\pm_n(x,t)$ per shell, with $|z_n^\pm|^2/2$ being the total energy in the shell $n$.
The equations read
\citep{Nigro_al_2004, Buchlin_Velli_2007}:
\begin{eqnarray}
k_{\perp n} = 2^n k_0 \ \ n=0... N \\
\partial_t z^\pm_n \pm B_0\partial_x z^\pm_n = T_n^\pm - \nu k_n^2 z^\pm_n + f^\pm_n
\label{shelleq}
\end{eqnarray}
where $f^+_n=f^-_n=f_n(x,t)$ represent the large scale (kinetic)
forcing, non zero for $n=0,1,2$ (see below), and the $T^\pm_n$ are the
nonlinear terms. These are made of a discrete sum of terms of the form:
$T^\pm_n = A k_{m}z^\mp_p z^\pm_q$ with $m$, $p$, $q$ being close to $n$, which
replace the convolution terms resulting from the Fourier transform with respect to the perpendicular coordinates of the original RMHD equations Eq.~\ref{s1}.
The control parameter is the time ratio $\chi_0$ imposed by the forcing term $f^\pm_n$:
\begin{equation}
\chi_0 = t_a^0/t_{NL}^0 = k_\perp^f b_{rms}/(k_\parallel^f B_0)
\label{chi0}
\end{equation}
where $k_\parallel^f$ and $k_\perp^f$ are the characteristic wave numbers of the forcing 
respectively in the parallel and perpendicular directions (note that $b_{rms} \simeq u_{rms}$).

\textit{Numerical method and parameters.}---We choose as in \citet{2008ApJ...672L..61P} the forcing correlation time
$t_{cor}^f$ to be smaller than the large scale Alfv\'en time
$t_a^0=(k^f_\parallel B_0)^{-1}$, $B_0=5$, and the aspect ratio of the
domain to be $L_\perp/L_z = 1/5$.
We define $K_0=2\pi/L_z$ and $k_0=2\pi/L_\perp$.
The smallest perpendicular forced wave number
is always $k_\perp^f = k_0$
(for a spectrum to develop, the Shell model needs to be excited at
least in three consecutive shells $k_0,~2k_0$, and $4k_0$).
Adding larger scales above forced scales doesn't modify the results.
We consider two forcings, varying $k_\parallel^f$: a narrow one or strong forcing ($k_\parallel^f=2 K_0$, $\chi_0=1$), and a wide one or weak forcing 
($k_\parallel^f\in[2,64] K_0$, $\chi_0=1/32$), estimating a priori in both cases
$b_{rms}=1$, $k_\perp^f \simeq 2$, and $k_\parallel^f
=\textrm{max}(k_\parallel^f)$ in Eq.~\ref{chi0}. 

Starting from the solutions $z^\pm_n(x,t)\equiv z^\pm(x,k_\perp,t)$ of
Eq.~\ref{shelleq} (hereafter we drop the index $n$ in $k_{\perp n}$), we define 
three time scales that will be
used to verify if turbulence satisfies the critical balance (CB) condition. 
We focus on the $z^+$ signal, and check that using $z^-$ leads to the same results. 
The correlation time $t_{cor}(k_\perp)$ is defined as
the full-width-half-maximum (FWHM) of the autocorrelation function in time
$A(z^+)_t$ 
of the signal $z^+(x,k_\perp,t)$, computed at each position x and then averaged
over the spatial domain (which is homogeneous).
The correlation length $L^\parallel_{cor}(k_\perp)$ is defined in a similar way, as the FWHM of the
autocorrelation function in space $A(z^+)_x$ computed at each time $t$
and then averaged on $25t_{NL}$ (statistically stationary time series). 
Definitions are summarized below, as well as the turbulence strength $\chi(k_\perp)$:
\begin{eqnarray}
t_{cor}(k_\perp) &=& \langle \textrm{FWHM}[A(z^+)_t]\rangle_x \label{t1}\\
t_a(k_\perp) &=&  \langle\textrm{FWHM}[A(z^+)_x]\rangle_t /B_0 \label{t2}
=L^\parallel_{cor}(k_\perp)/B_0\\
t_{NL}(k_\perp) &=& 1/(k_\perp z^-_{rms}(k_\perp)) \label{t3}\\
\chi(k_\perp) &=& t_a(k_\perp)/t_{NL}(k_\perp) \label{t4}
\label{tpar}
\end{eqnarray}
where  $\langle z\rangle_s=1/S\int z ds$ stands for the average, 
$A(z)_s=1/S\int z(s)z(s-s') ds'$ for the autocorrelation function, and 
$s=t,~x$ indicate the coordinate over which the average/correlation
function are computed (time and space respectively).
With the above definitions, the CB condition reads $\chi=1$ or $t_a=t_{NL}$.
Note that the quantity $t_a$ is the lifetime of a signal of coherence size $L_{cor}^\parallel$ propagating at
the Alfv\'en speed, which should also be equal to $t_{cor}$: 
this will be checked as well.

Finally we give the expressions for the energy spectra.
Denoting by $\hat z^\pm(k_\parallel,k_\perp,t)$ the Fourier
transforms with respect to $x$ of the signals $z^\pm$, 
we compute the 3D time-averaged spectra, $E^\pm_3$, 
the 1D reduced parallel and perpendicular spectra $E^\pm_{\bot,\parallel}$ as
\begin{eqnarray}
E_3^\pm(k_\parallel,k_\perp) &=& k_\perp^{-2} \langle|\hat
z^\pm(k_\parallel,k_\perp,t)|^2\rangle_{t} \\
E_\perp^\pm(k_\perp) &=&  
2\pi \int dk_\parallel k_\perp E_3^\pm (k_\parallel,k_\perp)
=k_\perp^{-1}\langle |z^\pm(x,k_\perp,t)|^2
\rangle_{x,~t}\qquad
\label{1dperp}\\
E_\parallel^\pm(k_\parallel) &=& 
2\pi \int dk_\perp k_\perp E_3^\pm(k_\parallel,k_\perp)
\label{1dpar}
\end{eqnarray}
Note that $E^\pm= \int E^\pm_3 d^3k = \int E^\pm_\perp
dk_\perp = \int E^\pm_\parallel dk_\parallel$. 

For further comparison, we define the theoretical 3D angular spectrum resulting from the strong cascade 
as  \citep{1995ApJ...438..763G}
\begin{equation}
E_{CB}(k_\parallel,k_\perp) = k_\perp^{-m-q-1} f(\chi)
\label{ECB}
\end{equation}
where $m$ is the slope of the 1D perpendicular spectrum and $q$ 
follows from the CB condition (assuming $m=5/3$ we have $q=2/3$).
The boundary of the excited
spectrum is given by the $\chi=1$ curve, its parallel extent is given by the CB
condition: 
$k_\parallel \propto k_\perp^{2/3}$. 
This is expressed by $f(|\chi|\ge1) \simeq 1$ and $f(|\chi|\le1)\simeq0$.
Applying the definitions Eqs.~(\ref{1dperp}-\ref{1dpar}) one obtains the familiar reduced spectra
$E_\perp^\pm\propto k_\perp^{5/3}$ and $E_\parallel^\pm\propto
k_\parallel^{-2}$.
Eq.~\ref{ECB} requires adopting a frame attached to the local propagation axis,
i.e., the local mean field. However here the propagation axis is always $B_0$,
in the absence of nonlocal terms in the shell model. This allows to consider eq.~\ref{ECB} to be valid in the absolute frame.

\begin{figure}[t]
\begin{center}
\includegraphics [width=0.98\linewidth]{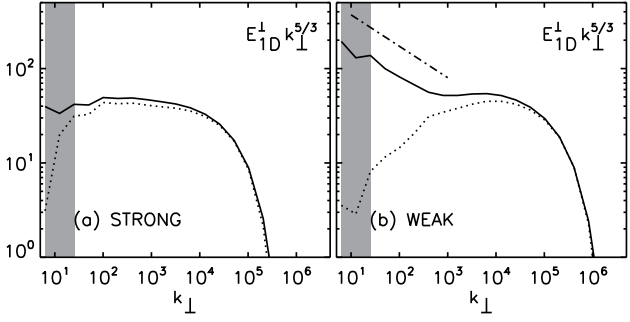}
\caption{(a): $\chi_0=1$. (b): $\chi_0=1/32$. Reduced {\it perpendicular} (total) energy spectra compensated by $k^{-5/3}$(solid lines). 
Perpendicular forcing scales are indicated as shaded areas in each panel. 
Dotted lines: reduced spectra built from the $E_3$ spectrum with $\chi<1/2$ excitation suppressed.
The dot-dashed line is the $k_\perp^{-2}$ scaling.}
\label{fig1}
\end{center}
\end{figure}

\textit{Results.}---In Fig.~\ref{fig1} we show the reduced {\it perpendicular} total energy
spectrum for the strong ($\chi_0=1$) and weak ($\chi_0=1/64$) forcing cases,
compensated by $k_\perp^{5/3}$. 
With strong forcing (left panel), the scaling is close to
$k_\perp^{-5/3}$ in the interval $100 \le k_\perp \le 10^4$.
With weak forcing (right panel) one sees two power laws, 
the scaling being
$k_\perp^{-2}$ at large scales and $k_\perp^{-5/3}$ at small scales.
Long integration times ($25 \ t_{NL}^0$) as well as large Reynolds number are needed to reveal this composite spectrum. 
We found equipartition between magnetic and kinetic energies in the weak
$k_\perp^{-2}$ range, while magnetic energy dominates by a uniform
factor $\simeq 2$ in the strong range (not shown).

\begin{figure}[t]
\begin{center}
\includegraphics [width=0.98\linewidth]{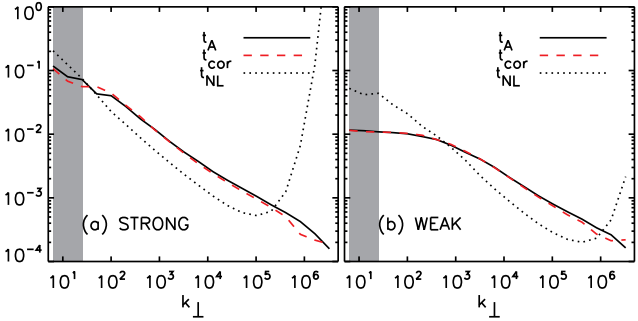}
\caption{(a): $\chi_0=1$. (b): $\chi_0=1/32$. Characteristic times $t_A,~t_{cor}$, and $t_{NL}$ for the signal
$z^+(x,k_\perp,t)$ as defined in Eqs.~\ref{t1}-\ref{t3}. Forcing scales are
marked in gray.} 
\label{fig2}
\end{center}
\end{figure}

We now examine the time scales defined in Eqs.~(\ref{t1}-\ref{t3}).
They are plotted versus $k_\perp$ in Fig.~\ref{fig2}.
Except at the forcing scales (gray band) and a bit below, the autocorrelation
and the Alfv\'en times (resp. dashed and solid lines) are superposed, showing coherence of the method.
In the strong forcing case (Fig.~\ref{fig2}a), the correlation time is
about twice the nonlinear time (dotted line) in the whole
inertial $-5/3$ range (CB condition). In the weak
forcing case (Fig.~\ref{fig2}b), the correlation time is 
constant at large scales, being given by the Alfv\'en time based on the parallel forcing scale. 
Then, at small scales, it switches to a value which is again about twice the nonlinear time.
This is compatible with the CB theory, which predicts the change in perpendicular spectral slope observed previously in Fig.~\ref{fig1}b, the CB condition
holding for $k_\perp\gtrsim10^3$. 

\begin{figure}[t]
\begin{center}
\includegraphics [width=0.98\linewidth]{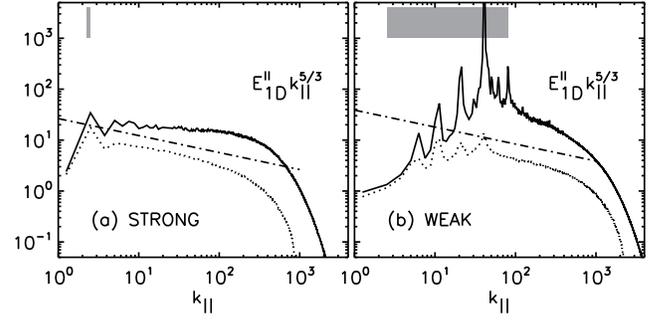}
\caption{(a): $\chi_0=1$. (b): $\chi_0=1/32$. Same format as Fig.~\ref{fig1} but for the reduced {\it parallel}
energy spectra. The dot-dashed line is the
$k_\parallel^{-2}$ scaling.}
\label{fig3}
\end{center}
\end{figure}

In Fig.~\ref{fig3} we show the reduced {\it parallel} total energy 
spectra, again for the strong and weak forcing cases (panel (a) and (b) respectively)
compensated by $k_\parallel^{5/3}$.
With strong forcing, the parallel spectrum scales as $\simeq k_\parallel^{-1.8}$. Hence it is steeper than the perpendicular spectrum but flatter than the CB
prediction (slope -2). With weak forcing the spectrum at low
wave numbers is shaped by the forced parallel modes (gray band on
top). Their signature persists because the cascade is weak there ($t_A\ll
t_{NL}$ in Fig.~\ref{fig2}b). 
For higher wave numbers one expects to find the strong cascade scaling
as in panel (a), on the contrary the slope is steeper, even steeper
than the -2 predicted by the CB phenomenology.

\begin{figure}[t]
\begin{center}
\includegraphics [width=0.49\linewidth]{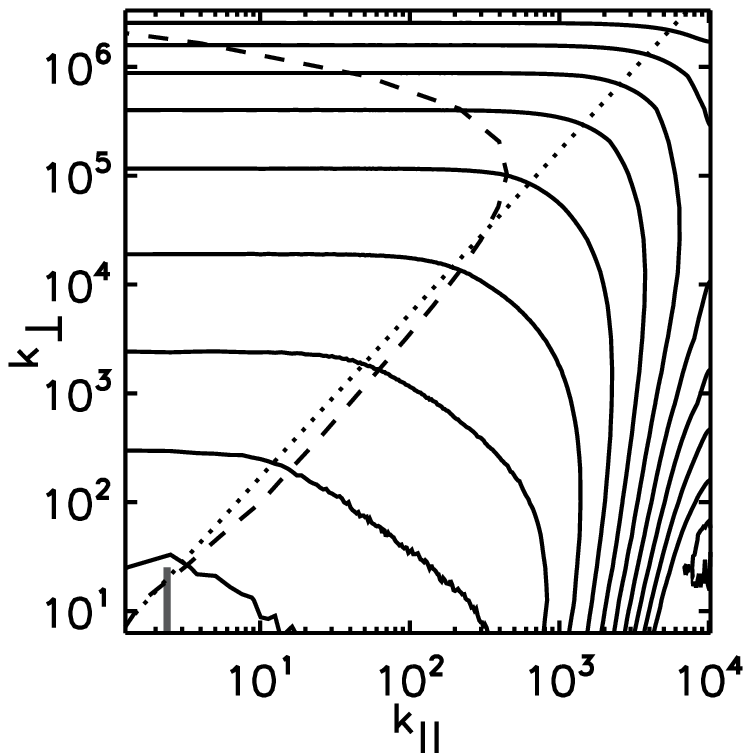}
\includegraphics [width=0.49\linewidth]{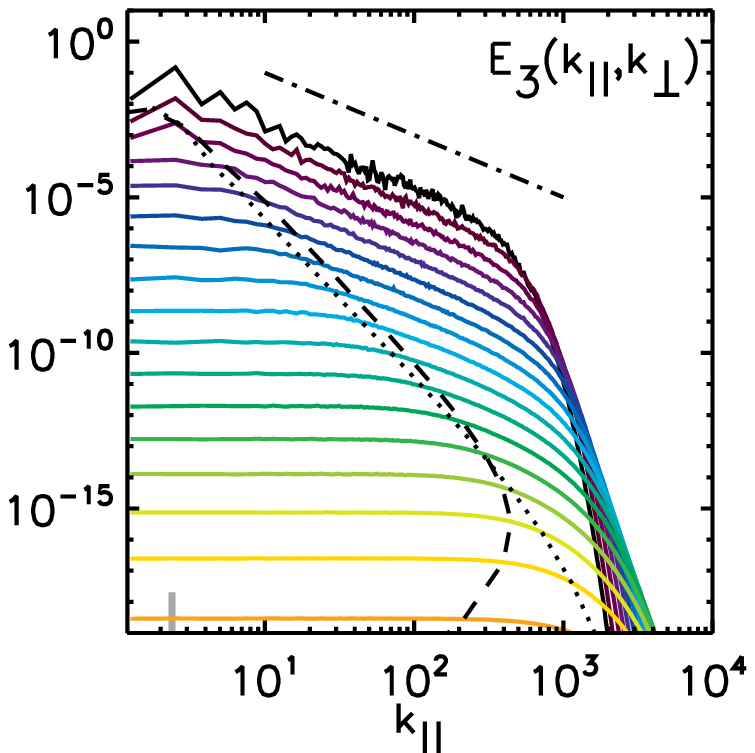}
\caption{
Anisotropy properties ($\chi_0=1$).
Left panel: contours of energy spectrum $E_3$ in the $(k_\parallel,k_\perp)$ plane. 
Right panel: horizontal cuts of $E_3$ vs $k_\parallel$ for $k_\perp = k_0 \
2^n$ with $n=0$ to $16$ from top to bottom. 
Dot-dashed line: $k_\parallel^{-2}$ scaling. In both panels the dotted line is the theoretical boundary $\chi=1$;
the dashed line is the measured boundary $\chi=1$. Forced scales are marked by
a shaded area.}
\label{fig4}
\end{center}
\end{figure}

The different parallel slopes found in the reduced spectra can be understood
by considering the underlying \textit{3D angular spectrum}. 
We first examine the strong forcing case, Fig.~\ref{fig4}. Isocontours of
time-averaged 3D spectrum $E_3(k_\parallel,k_\perp)$ are plotted in the left
panel.
We give two representations of the CB:
a theoretical $\chi=1$ contour, adopting strictly the scaling $k_\parallel
\propto k_\perp^{2/3}$ (dotted line) and an effective $\chi=1$ contour, using the computed $b_{rms}(k_\perp)$ (dashed line). They differ significantly
only when entering in the perpendicular dissipation range.
The CB condition, $\chi=1$, separates two regions:
strong coupling is found on the left of the $\chi=1$ line, 
weak coupling on the right. 
One can see that the energy isocontours are mostly horizontal in the strong region $\chi>1$;
the spacing between successive contours corresponds nearly to the scaling
$k_\perp^{-10/3}$ (Eq.~\ref{ECB}), in the range $10^2<k_\perp <10^4$.
A remarkable feature is that, contrary to the usual assumption, the energy
density doesn't drop abruptly when entering the weak coupling region
($\chi<1$), at least for large perpendicular scales.
This is best seen in the horizontal cuts (right panel of Fig.~\ref{fig4}),
where again the CB is overplotted. For $E_3\gtrsim10^{-13}$
(corresponding to $k_\perp\lesssim 10^4$),
the energy density $E_3$ decreases as $k_\parallel^{-2}$. The extent of this
scaling decreases progressively with increasing $k_\perp$.
In the CB phenomenology, the reduced parallel scaling in $k_\parallel^{-2}$ is
due to the contribution of parallel wave numbers satisfying $\chi=1$, which
occurs only at small perpendicular scales
since parallel excitation is assumed to be negligible for $\chi <1$.
Here, on the contrary, the dominant contribution to the reduced parallel scaling is due to the large perpendicular scales (with $\chi \ll 1$).
This is demonstrated in Fig.~\ref{fig3}a where the dotted line, with the
typical scaling $k_\parallel^{-2}$, represents the equivalent CB
spectrum, 
i.e. the reduced parallel spectrum obtained from $E_3$ 
after suppressing excitation in the region $\chi<1/2$: one sees that the resulting spectrum is much closer to the $k_\parallel^{-2}$ scaling than the full solution.

\begin{figure}[t]
\begin{center}
\includegraphics [width=0.49\linewidth]{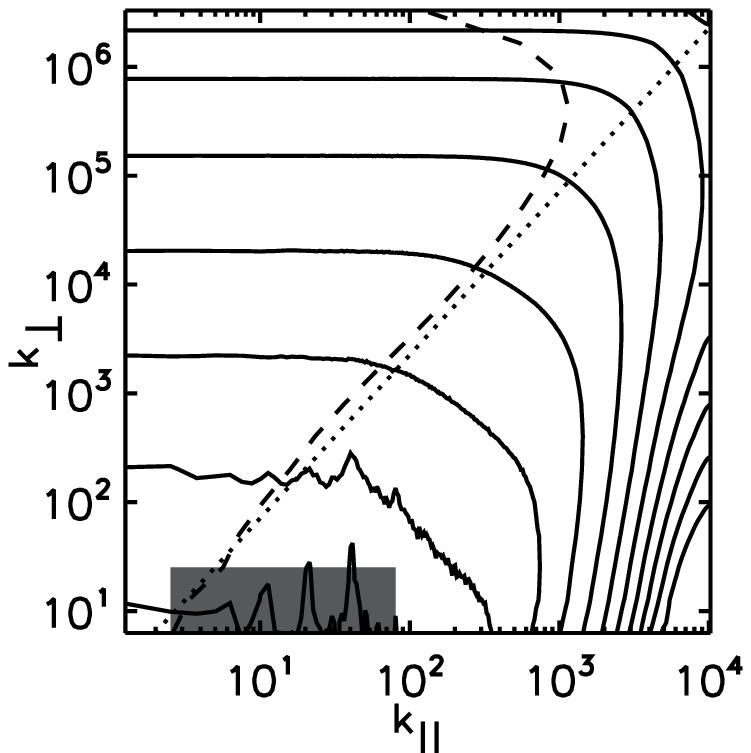}
\includegraphics [width=0.49\linewidth]{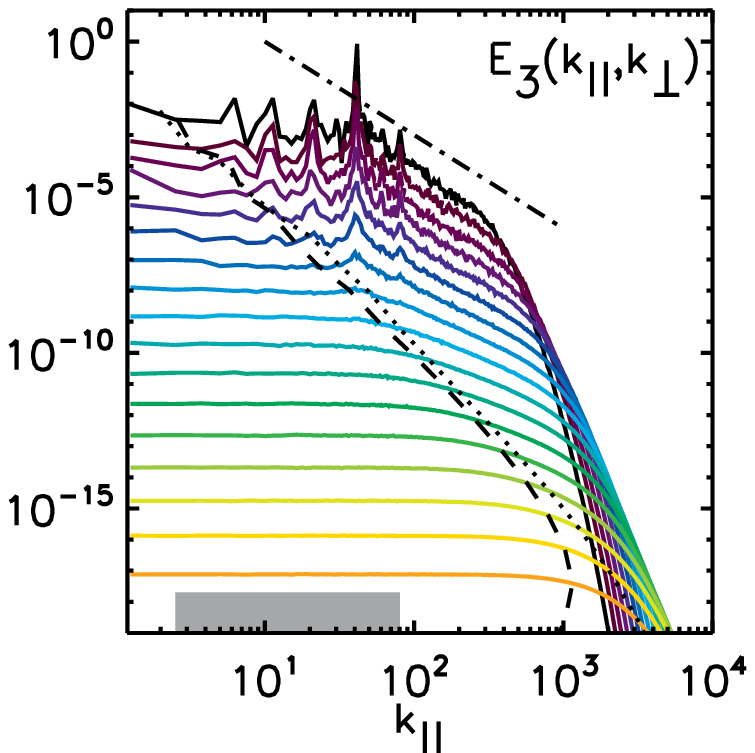}
\caption{Anisotropy properties ($\chi_0=1/32$).
Same caption as Fig.~\ref{fig4}. The dot-dashed line 
is the $k_\parallel^{-3}$ scaling.}
\label{fig5}
\end{center}
\end{figure}

We consider now the weak forcing case in Fig.~\ref{fig5}. The
energy isoncontours for $k_\perp<10^3$ reveal a cascade at constant parallel
wave number ($k_\parallel \lesssim 10^2$).
Then, at higher $k_\perp$, the parallel spectrum widens according to the scaling
$k_\parallel\propto k_\perp^{2/3}$.
Horizontal cuts of the 3D spectrum (right panel) have features similar to the
strong forcing case: an inertial strong cascade and a
steep spectrum at intermediate and large perpendicular scales respectively,
but now shifted to the last parallel
forced mode ($k_\parallel^f\simeq 10^2$).
Again there is no abrupt energy decrease when passing the $\chi=1$ boundary,
however the scaling is now $\propto k_\parallel^{-3}$, thus explaining the
steeper spectral slope found in the reduced parallel spectrum
in Fig.~\ref{fig3}b. 

\textit{Discussion.}---
In Reduced MHD, parallel structures reflect directly the temporal structure of
the perpendicular nonlinear excitations via the linear propagation of Alfv\'en waves.
Hence, the $k_\parallel^{-2}$ scaling of the large perpendicular
scales should result from a $f^{-2}$ spectrum of the low perpendicular wave numbers
(in the weak case, the slope $-3$ plays the same role).
We have checked that the $f^{-2}$ spectrum is indeed present (see also
\citep{Buchlin_Velli_2007,Nigro_al_2008, Verdini_al_2012,2007ApJ...657L..47R}).
This scaling is not due to forcing since
suppressing nonlinear couplings leads to a
steeper spectrum (slope $k_\parallel^{-4}$). 
The $k_\parallel^{-2}$ scaling develops in the unforced case after a few nonlinear times and lasts for more than ten nonlinear times. 
We conjecture that the high frequencies in large eddies are due to
an \textit{inverse transfer} from small to large scales, which may occur even in absence of a proper inverse cascade associated with a
definite invariant \citep{Gloaguen_al_1985}.

Critical balance tests gave good results, with $t_{cor}$ about twice $t_{NL}$ in the strong inertial range (Fig.~\ref{fig2}). 
We have found that, although globally balanced, the two Alfv\'en species spectral amplitudes show also a local imbalance of a factor $3$. This might explain the factor two in time scales, which as well may be due to arbitrariness in defining the nonlinear time.
It is worthwhile noting that reduced spectra in the solar
wind agree with the form dictated by the CB condition,
but that an excess of parallel energy may be required \citep{Forman_al_2011}, consistent with our findings.
Much is to be gained by testing turbulence theories using the Shell-RMHD
model, especially with respect to the two points just mentioned:
the conditions of
appearance for the local scaling (local $z^+/z^-$ imbalance) which controls the
global spectral scaling and the origin of the frequency spectrum of large eddies which controls the anisotropy of turbulence.


We thank W.-C. M\"uller, M. Velli, and \"O. G\"urcan for useful
discussions. A.V. acknowledges support from the Belgian Federal Science Policy Office (ESA-PRODEX program).


\end{document}